\documentclass[12pt]{article}

\usepackage{amsmath,amssymb,graphicx} 
\usepackage{epic}
\usepackage{epsfig}

\newcommand{\beq}{\begin{eqnarray}}
\newcommand{\eeq}{\end{eqnarray}}
\newcommand{\refcite}{\cite}

\begin{document}

\title{\bf Continuum Superpartners\vspace{0.3in}}

\author{\textbf{Hsin-Chia Cheng}\vspace{0.2in}
\\
\normalsize{\it Department of Physics, University of California}\\
\normalsize{\it Davis, California 95616, USA}}

\date{}

\maketitle
\vspace{0.5in}

\begin{abstract}
In an exact conformal theory there is no particle. The excitations have continuum spectra and are called ``unparticles'' by Georgi. We consider supersymmetric extensions of the Standard Model with approximate conformal sectors.  The conformal symmetry is softly broken in the infrared which generates a gap. However, the spectrum can still have a continuum above the gap if there is no confinement. Using the AdS/CFT correspondence this can be achieved with a soft wall in the warped extra dimension. When supersymmetry is broken the superpartners of the Standard Model particles may simply be a continuum above gap. The collider signals can be quite different from the standard supersymmetric scenarios and the experimental searches for the continuum superpartners can be very challenging.
\end{abstract}

\vspace{0.8in}
\noindent {\it Talk presented at 2009 Nagoya Global COE Workshop -- Strong Coupling Gauge Theories in LHC Era (SCGT09), 8--11 December, 2009}


\newpage

\section{Introduction}

The Large Hadron Collider (LHC) has started running. With its unprecedented center of mass energy, there is a high hope that it will discover new physics which revolutionizes high energy physics. However, LHC is a complicated machine with very high luminosities. In order to find new physics, we need to know what signals the new physics may give rise to, and how to search for them with the enormous Standard Model (SM) backgrounds. 

A major motivation for new physics at the TeV scale is the hierarchy problem -- the stability of the electroweak scale under radiative corrections. Many models for TeV-scale new physics are invented to address this problem, {\it e.g.,} supersymmetry (SUSY), technicolor models, large and warped extra dimensions, little Higgs models, etc. A lot of collider phenomenology studies have been devoted to these models. However, these models certainly do not cover all possible new physics that may appear at the LHC. There is no theorem that at the TeV scale we should only see the minimal models which just address the hierarchy problem.

{}From the experimental point of view, the collider searches should be signal-based. Even though there are experimental studies targeted for some specific models, generic searches for new particles such as new gauge bosons, new quarks and leptons apply to a wide range of models.  As the LHC is starting, the recent model-building efforts have shifted from ``solving problems of the Standard Model'' to studying models which can give rise to ``unexpected signals'' ({\it e.g.,} hidden valleys~\cite{Strassler:2006im}, quirks~\cite{Kang:2008ea}, unparticles~\cite{Georgi:2007ek}, etc.) irrespective of whether they are related to any particular problem of the Standard Model. It is possible that the TeV-scale physics includes some of these extra sectors in additional to the standard scenario. The presence of the extra sector may obscure the experimental signals of the standard scenario if there is mixing between them, so it is imperative to study these possibilities and their experimental consequences. In this talk, consider such a case with a (super)conformal sector softly broken at the TeV scale and mixed with the supersymmetric Standard Model. As we will see, the mixings can make the superpartners of the SM particles have continuum spectra, and hence make their experimental consequences quite different from the standard SUSY scenario. This presentation is based on the work done in collaboration with Haiying Cai, Anibal Medina, and John Terning~\cite{Cai:2009ax}.

\section{Unparticles}

We first give a brief introduction to unparticles. In Ref.~\refcite{Georgi:2007ek} Georgi considered the possibility that a hidden conformal field theory (CFT) sector coupled to the SM through higher-dimensional operators,
\begin{equation} 
\frac{C_U \Lambda_U^{d_{UV}-d_U}}{M_U^k} {\cal O}_{SM} {\cal O}_U,
\end{equation}
where $M_U$ is the scale where the interaction is induced, $\Lambda_U$ is the scale where the hidden sector becomes conformal, $d_{UV}= k+4 - d ({\cal O}_{SM})$, and $d_U$ is the scaling dimension of the operator ${\cal O}_U$.
There is no ``particle''  in a CFT. One can only talk about operators with some scaling dimensions. The spectral densities are continuous and hence Georgi called them ``unparticles.''

The phase space of an unparticle looks like a fractional number of particles. For a scalar unparticle,
\begin{eqnarray}
\langle {\cal O}_U (x) {\cal O}_U (0)\rangle = \int \frac{d^4 P}{(2\pi)^4} e^{-i Px } |\langle 0 | {\cal O}_U | P\rangle |^2 \rho (P^2) , \nonumber \\
 |\langle 0 | {\cal O}_U | P\rangle |^2 \rho (P^2) = A_{d_U} \theta(P^0) \theta (P^2) (P^2)^{d_U -2},
\end{eqnarray}
where $A_{d_U}$ is a normalization constant. In the limit $d_U \to 1$, it becomes the phase space of a single massless particle, $\delta (P^2)$. For $d_U \neq 1$, the spectral density is continuous.

Because the unparticle spectral density is continuous down to zero, many phenomenological consequences and constraints are similar to other scenarios with very light (or massless) degrees of freedom, {\it e.g.,} large extra dimensions proposed by Arkani-Hamed, Dimopoulos, and Dvali~\cite{ArkaniHamed:1998rs}. A list of things to be considered contains\footnote{There is a huge literature on phenomenology tests and constraints on unparticles. We only list a few sample references due to the space limit. Please check the citation list of the original Gerogi's paper for more complete references.}
\begin{itemize}
\item invisible decays, {\it e.g.,} from $Z$, heavy mesons, quarkonia, neutrinos, \ldots~\cite{Aliev:2007gr,Chen:2007zy},
\item missing energy in high energy collisions~\cite{Cheung:2007zza},
\item higher-dimensional operators induced by unparticles~\cite{Cheung:2007zza,Georgi:2007si,Luo:2007bq},
\item astrophysics: star (SN1987A) cooling~\cite{Davoudiasl:2007jr,Hannestad:2007ys},
\item cosmology: Big Bang nucleosynthesis (BBN)~\cite{Davoudiasl:2007jr,McDonald:2007bt},
\item long-range forces induced by unparticles~\cite{Liao:2007ic,Goldberg:2007tt,Deshpande:2007mf}.
\end{itemize}
In particular, unparticles cannot carry SM charges if the spectral density continues down to zero without any gap. Otherwise they would have been copiously produced and observed.

However, as pointed by Fox, Rajaraman, and Shirman~\cite{Fox:2007sy}, coupling to the Higgs sector will break the conformal invariance of the unparticles in the infrared (IR), which will induce a gap in general. Above the gap, there are two possibilities:
\begin{itemize}
\item If the theory confines, then it would produce QCD-like resonances.
\item If the theory does not confine, one then expect that the spectral density is still continuous above the mass gap. A simple ansatz to describe it is to introduce an IR cutoff $m$~\cite{Fox:2007sy,Cacciapaglia:2007jq},
\begin{eqnarray}
\Delta (p, m, d) & \equiv & \int d^4 x e^{ipx} \langle 0 | T {\cal O} (x) {\cal O}^\dagger (0) |0\rangle \nonumber \\
&=& \frac{A_d}{2\pi} \int_{m^2}^{\infty} (M^2 -m^2)^{d-2} \frac{i}{p^2 - M^2 + i \epsilon } d M^2 .
\end{eqnarray}
Such a spectral density arises when a massive particle couples to massless degrees of freedom. For example, quark jets can be considered as unparticles~\cite{Neubert:2007kh}.
\end{itemize}

With a mass gap, many low-energy phenomenological constraints which result from new light degrees of freedom no longer apply. In particular, with a mass gap, one can consider the possibility that unparticles carry SM charges~\cite{Cacciapaglia:2007jq} (if the gap is larger than ${\cal O}(100)$ GeV), which can be more interesting phenomenologically. This is the scenario that will be discussed in this talk.

\section{Unparticles and AdS/CFT}

A useful tool to study the (large $N$) conformal field theory is the AdS/CFT correspondence~\cite{Maldacena:1997re}. The metric of a 5D anti de Sitter (AdS$_5$) space in conformal coordinates is given by
\begin{equation}
ds^2 = \frac{R^2}{z^2} (dx_\mu^2 - dz^2),
\end{equation}
where $R$ is the curvature radius and $z$ is the coordinate in the extra dimension.
Using the AdS/CFT correspondence, Georgi's unparticle scenario can be described by the Randall-Sundrum II (RS2)~\cite{Randall:1999vf} setup. The 5-dimensional (5D) bulk fields correspond to the 4-dimensional (4D) CFT operators, and the 5D bulk mass of a bulk field is related to the scaling dimension of the corresponding CFT operator. In RS2, the space is cut off by a UV brane (located at $z=z_{UV}\equiv \epsilon$) but there is no IR brane which means that the conformal invariance is good down to zero energy. The momentum in the extra dimension of a bulk field is not quantized.  From the 4D point of view, the momentum in the extra dimension becomes the mass in 4D, so the 4D mass spectrum is continuous. SM fields are localized on the UV brane. They can interact with the bulk fields (CFT operators) through higher-dimensional interactions. Such a representation allows us to perform calculations involving unparticles using the ordinary (5D) field theory~\cite{Cacciapaglia:2008ns,Falkowski:2008yr,Friedland:2009iy}.

In this talk we are interested in unparticles carrying SM charges with mass gaps. Some modifications of the RS2 setup are necessary. In particular, SM fields have to propagate in the extra dimension as well and a soft wall is needed to produce the gap in the IR. Before going to these discussions, we first describe how to formulate SUSY in the AdS bulk.

\subsection{SUSY in AdS bulk}

It is convenient to describe a 5D supersymmetric theory using the 4D $N=1$ superspace formalism~\cite{ArkaniHamed:2001tb,Marti:2001iw}. The 5D $N=1$ SUSY corresponds to $N=2$ SUSY in 4D. A 5D $N=1$ hypermultiplet contains two chiral multiplets in 4D, $\Phi= \{\phi, \chi, F\},\, \Phi_c = \{\phi_c, \psi, F_c\}$. The action of the 5D $N=1$ hypermultiplet is given by~\cite{Marti:2001iw}
\begin{eqnarray}
S & =& \int d^4 x\, d z \left\{
\int d^4 \theta\, \left( \frac{R}{z} \right)^3
\left[ \Phi^\ast\, \Phi + \Phi_c\, \Phi_c^\ast\right] + \right.\nonumber\\
& & \left.+ \int d^2 \theta\, \left( \frac{R}{z} \right)^3  \left[
\frac{1}{2}\; \Phi_c\, \partial_z \Phi - \frac{1}{2}\; \partial_z
\Phi_c\, \Phi + m \frac{R}{z}\; \Phi_c\, \Phi \right] + h.c.
\right\}\,,
\label{5Daction} 
\end{eqnarray}
It is convenient to define a dimensionless bulk mass $c \equiv mR$. It is related to the dimension of the corresponding operator in the 4D CFT picture~\cite{Aharony:1999ti,Contino:2004vy,Cacciapaglia:2008bi}. For a left-handed CFT operator ${\cal O}_L$ (which corresponds to the 5D bulk field $\Phi$), we have $d_s = 3/2 -c$, $d_f =2-c$ for $c \leq 1/2$, where $d_s$ and $d_f$ are the scaling dimensions of the scalar and fermion operators respectively. For $c <-1/2$ ($d_s>2$), the correlator diverges as we take the UV brane to the AdS boundary ($\epsilon \to 0$). A counter term is needed on the UV brane to cancel the divergence, which implies UV sensitivity. On the other hand, for $c> 1/2$ the CFT becomes a free field theory with $d_s=1$, $d_f=3/2$. We will focus on the most interesting range $-1/2 \leq c \leq 1/2$ ($ 1\leq d_s \leq 2$). For the right-handed CFT operator, we have the similar interpretations but with $c \to -c$.

In solving the equations of motion (EOMs), we include a $z$-dependent mass $m(z)$ which represents the soft wall. The EOMs for the fermions and the $F$-terms are
\begin{eqnarray}
 - i\bar \sigma ^\mu  \partial _\mu \chi  - \partial _z \bar \psi  + (m(z)R + 2)\frac{1}{z} \bar \psi  &=& 0 ,\label{fermion1}\\
- i\sigma ^\mu  \partial _\mu  \bar \psi  + \partial _z \chi  +
(m(z)R - 2)\frac{1}{z}\chi  &=& 0\label{fermion2}.
\end{eqnarray}
\begin{eqnarray}
F_c^ * &=&  {-\partial _z \phi + \left(\frac{3}{2} - m(z)R\right)\frac{1}{z}} \phi \label{fc} ,\\
F &=&  {\partial _z \phi_c^* -\left(\frac{3}{2} +m(z)R\right)\frac{1}{z}}
\phi^*_c .
\end{eqnarray}
Using the $F$-term equations we can find the second order EOM for the scalars:
\begin{eqnarray}
\!\!\!\! \partial_\mu \partial^\mu \phi - \partial_z^2 \phi + \frac{3}{z}\; \partial_z \phi + \left( m(z)^2 R^2+ m(z)R - \frac{15}{4} \right) \frac{1}{z^2}\; \phi  -\left(\partial_z m(z)\right) \frac{R}{z}\phi=0\,.
\label{scalar} 
\end{eqnarray}
We can decompose the 5D field into a product of the 4D field and a profile in the extra dimensions
\begin{eqnarray}
 \chi (p,z) & = &  \chi _{\rm{4}} (p)\left(\frac{z}{z_{UV}}\right)^2 f_L(p,z),  \quad  \phi (p,z) =  \phi _{\rm{4}} (p)\left(\frac{z}{z_{UV}}\right)^{3/2} f_L(p,z), \label{decom1}\\
 \psi (p,z) & = &  \psi _{\rm{4}} (p)\left(\frac{z}{z_{UV}}\right)^2 f_R(p,z), \quad
\phi _c (p,z) =  \phi _{{\rm{c4}}}
(p)\left(\frac{z}{z_{UV}}\right)^{3/2} f_R(p,z), \label{decom2}
 \end{eqnarray}
where $p \equiv \sqrt{p^2}$ represents the 4D momentum.
Then we find the profiles in the extra dimension $f_L, f_R$ satisfy the Schr\"{o}dinger-like equations with the potential determined by the mass term. For a constant mass, the potential scales as $1/z^2$ for large $z$, so there is a continuum of solutions starting from zero energy (4D mass). We can get different behaviors if $m \sim z^\alpha$ for large $z$:
\begin{itemize}
\item For $\alpha < 1$, the potential still goes to zero for large $z$, so we have continuum solutions without gap.
\item For $\alpha > 1$, the potential grows without bounds at large $z$, so we get discrete solutions~\cite{Karch:2006pv}.
\item For $\alpha =1$, the potential approaches a positive constant for large $z$, so we have a continuum with a gap~\cite{Cacciapaglia:2008ns}.
\end{itemize}
We are interested in the last possibility so we consider the case $m(z)=c+\mu z$. In the UV (small $z$), $m \approx c$, we have the usual CFT interpretation. At large $z$, $m(z) \propto z$, the conformal invariance in broken in the IR and a mass gap is developed.

We can easily obtain the zero mode solutions for $m(z)=c+\mu z$:
\begin{equation}
f_L (0, z) \sim e^{-\mu z} z^{-c}, \qquad f_R (0, z) \sim e^{\mu z} z^c .
\end{equation}
One can see that only one of them has a normalizable zero mode. For definiteness we choose $\mu >0$, then only the left-handed field has a zero mode, which is identified as the SM fermion.

The nonzero modes satisfy the Schr\"{o}dinger-like equations,
\begin{eqnarray}
\frac{\partial^2}{\partial z^2}f_R+\left(p^2-\mu^2-2\frac{\mu
c}{z}-\frac{c(c-1)}{z^2}\right)f_R&=&0\label{2ndorder1},\\
\frac{\partial^2}{\partial z^2}f_L+\left(p^2-\mu^2-2\frac{\mu
c}{z}-\frac{c(c+1)}{z^2}\right)f_L&=&0 \label{2ndorder2}.
\end{eqnarray}
They have the same form as the radial wave equation of the hydrogen atom,
\begin{eqnarray}
\frac{\partial^2}{\partial r^2}u+\left(2 M E+2\frac{M
\alpha}{r}-\frac{\ell(\ell+1)}{r^2}\right)u&=&0 ,
\end{eqnarray}
except that $c$ is in general a fractional number instead of an integer. We can immediately see the pattern of the solutions from our knowledge of the hydrogen atom. For $c\mu >0$ which corresponds to a repulsive Coulomb potential, we have continuum solutions above the gap, $p^2 > \mu^2$. For $c\mu < 0$ on the other hand, we get discrete hydrogen-like spectrum below the gap in addition to the continuum above the gap.

\subsection{Holographic Boundary Action}

One can derive the unparticle propagator by probing the CFT with a boundary source field $\Phi_c^0$. The holographic boundary action can be obtained by integrating out the bulk using the bulk EOMs with the boundary condition $\Phi_c (z_{UV}) = \Phi_c^0 = (\phi_c^0, \psi^0, F_c^0)$~\cite{Cacciapaglia:2008ns},%
\begin{eqnarray}
S_{holo}  =  - \int {d^4 } x[\phi _{c}^{0*}  \Sigma_{\phi_c} \phi
_{c}^{0}  + F_{c}^{0*}  \Sigma_{F_c} F_{c0}  + \psi _{0}^ *
\Sigma_{\psi} \psi _{0} ]\label{holoaction}
\end{eqnarray}
where
\begin{equation}
\Sigma_{\phi_c}=\left(\frac{R}{z_{UV}}\right)^3p\frac{f_L}{f_R},\quad
\Sigma_{\psi}=\left(\frac{R}{z_{UV}}\right)^4\frac{p_\mu\sigma^\mu}{p}\frac{f_L}{f_R},\quad
\Sigma_{F_c}=\left(\frac{R}{z_{UV}}\right)^3\frac{1}{p}\frac{f_L}{f_R}.
\end{equation}
{}From the CFT point of view, the right-handed superfield $\Phi_c^0$ is  a source of the left-handed CFT operator which correspond to $\Phi$. Since $F_c^0$ is the source for the scalar component, the propagator for the scalar CFT operator is 
\begin{equation}
\Delta_s (p) \propto - \Sigma_{F_c} (p),
\end{equation} 
with
\begin{eqnarray}
\Sigma_{F_c} =  \frac{\epsilon(\mu +\sqrt {\mu^2  - p^2 })}{p^2
}\cdot \frac{{W\left( { - \frac{{c\mu}}{{\sqrt {  \mu^2 - p^2 }
}},\frac{1}{2} + c,2\sqrt { \mu^2 - p^2  } \epsilon}
\right)}}{{W\left( { - \frac{{c\mu}}{{\sqrt {  \mu^2 - p^2 }
}},\frac{1}{2} - c,2\sqrt {  \mu^2 - p^2 } \epsilon} \right)}},
\end{eqnarray}
where $W(\kappa, m, \zeta)$ is the Whittaker function of the second kind.
The fermionic and the $F$-component CFT correlators are simply $\Delta_f = p_\mu \sigma^\mu \Delta_s$ and $\Delta_F = p^2 \Delta_s$ by SUSY relations.

We are interested in the conformal limit $\epsilon \to 0$, so we expand $\Sigma_{F_c}$ for small $\epsilon$ and focus on the range $-1/2 < c < 1/2$. After properly rescaling the correlator by a power of $\epsilon$ to account for the correct dimension of the correlator, we find
\begin{equation}
\Delta_{s}^{-1}\approx { \frac{{p^2 }}{{1 - 2c}}\epsilon^{1-2c} - \frac{{2^{ - 1 + 2c} p^2 (\mu^2  - p^2 )^{ - 1/2 + c}
\Gamma (1 - 2c)\Gamma (c + \frac{{c\mu}}{{\sqrt {\mu^2  - p^2 }
}})}}{{\Gamma (2c)\Gamma (1 - c + \frac{{c\mu}}{{\sqrt {\mu^2  -
p^2 } }})}}}. \label{delta}
\end{equation}
The first term vanishes in the conformal limit for $-1/2 < c < 1/2$. We can see that there is a pole at $p^2=0$ which represents the zero mode, and a branch cut for $p^2 >\mu^2$ which represents the continuum in the propagator. For $c<0$, there are addition poles below the gap $\mu^2$ which correspond to the hydrogen-like bound states discussed in the previous subsection.

\subsection{SUSY Breaking}

Now we introduce SUSY breaking on the UV boundary by a scalar mass term,
\begin{equation}
\delta S = \frac{1}{2} \int {d^4 x}\left( \frac{R}{z_{UV}} \right
)^3 \int {dz\left( {m^2 z_{UV} \cdot \phi ^ *  \phi  + h.c.}
\right)} \delta (z - z_{UV})\, .\label{boundarysusybreaking}
\end{equation}
The scalar propagator is modified due to the modified boundary conditions,
\begin{equation}
F_c(z_{UV} ) = F_{c}^{0}  + m^2 z_{UV}\, \phi^ * (z_{UV}) , \qquad
\psi (z_{UV} ) = \psi^0 , \qquad \phi_c (z_{UV} ) = \phi _{c}^{0}
\end{equation}
One can repeat the analysis of the previous subsection and finds for $-1/2 < c < 1/2$ in the limit $\epsilon \to 0$,
\begin{equation}
\Delta_{s}^{-1}(p^2) \approx  m^2 \epsilon ^{  1
-2c}-\frac{p^2}{2c-1}\epsilon^{1-2c} - \frac{2^{ - 1 + 2c} p^2
(\mu^2 - p^2 )^{ - 1/2 + c} \Gamma (1 - 2c)\Gamma (c +
\frac{c\mu}{\sqrt {\mu^2 - p^2 }})}{\Gamma (2c)\Gamma (1 - c +
\frac{c\mu}{\sqrt {\mu^2 - p^2 } })}.
\label{susybreaking2}
\end{equation}
For small $m^2$, the pole which was at zero mass is shifted to
\begin{equation}
p^2_{pole}=\frac{2(2c-1)m^2(\mu\epsilon)^{1-2c}}{(-4^{c}+2^{1+2c}c)\Gamma(1-2c)}. \label{displacedpole}
\end{equation}
\begin{figure}
\begin{center}
\epsfig{file=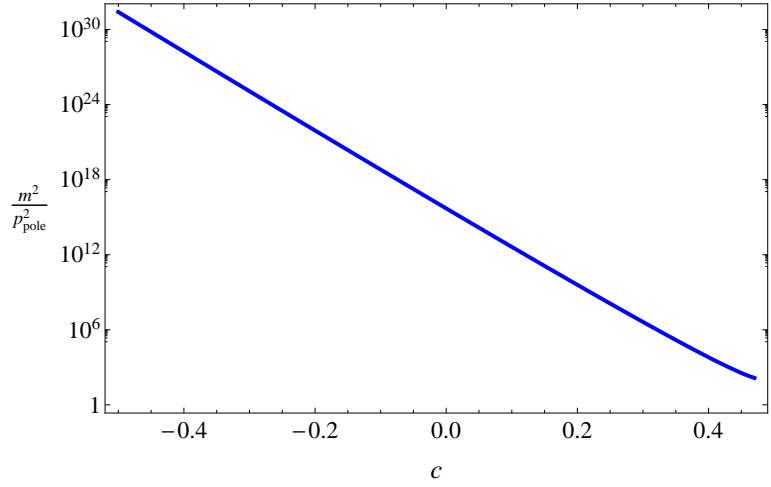,width=4.0in}
\end{center}
\caption{Plot of $m^2/p^2_{pole}$ vs $c$. Notice that as $c$ gets
closer to $1/2$, for a given value of $p^2_{pole}$, the value of
$m^2$ decreases. }
\label{logm2}
\end{figure}
The shift is smaller for smaller (more negative) $c$ (Fig.~\ref{logm2}), because the zero mode wave function is localized farther away from the UV brane ($f_{L,0} \propto e^{-\mu z} z^{-c}$).

\begin{figure}[H]
\begin{center}
\epsfig{file=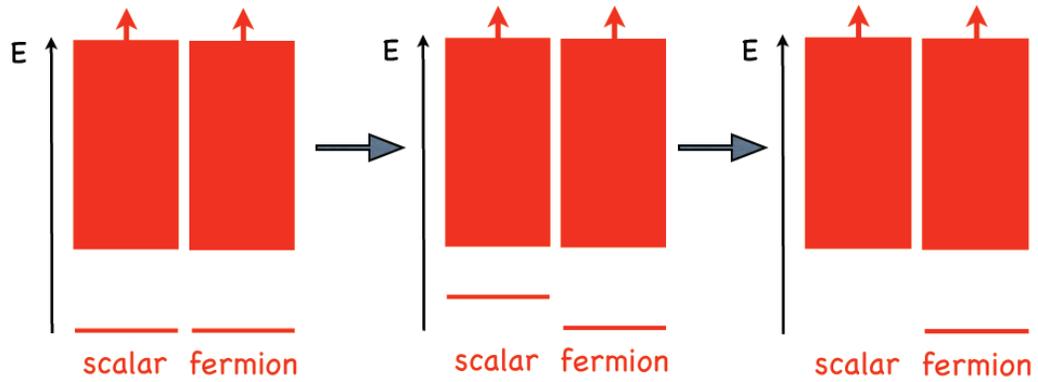,width=5.5in}
\end{center}
\caption{The spectra of the scalar and the fermion when we increase the scalar SUSY-breaking mass term on the UV brane.}
\label{merge}
\end{figure}

\begin{figure}[t]
\begin{center}
\epsfig{file=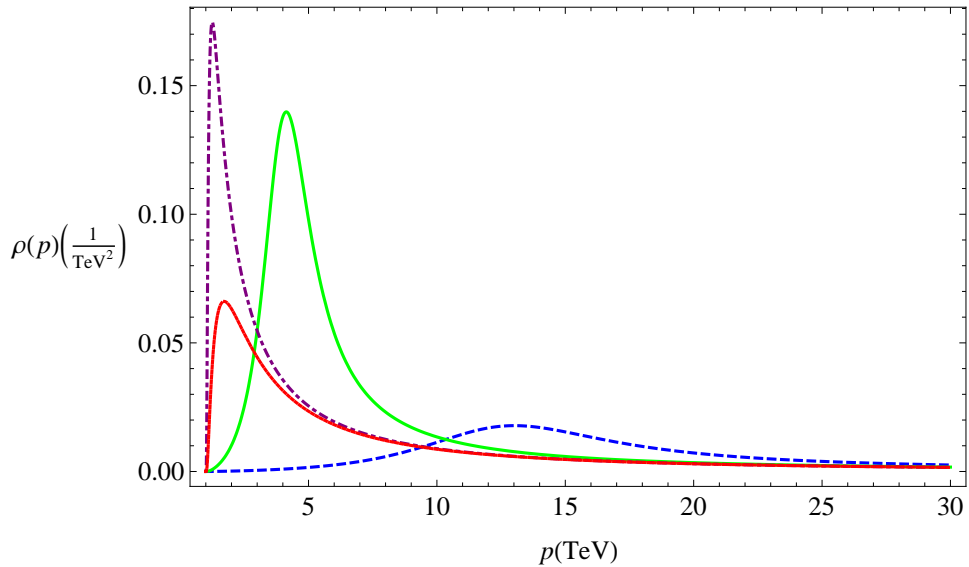,width=5.0in}
\end{center}
\caption{Spectral function for four examples of boundary UV SUSY
breaking masses $m=2\times10^{7}$ GeV (blue curve, dashed),
$m=8\times 10^{6}$ GeV (green curve, solid), $m=2\times 10^{6}$
GeV (purple curve, dot-dashed) and $m= 10^{5}$ GeV (red curve,
dotted). The red and purple curves correspond to zero-mode poles
localized at $p\approx 50$ GeV (red curve) and $p\approx 950$ GeV
(purple curve), that haven't merged with the continuum. On the
other hand, the green and blue curves correspond to the cases
where the pole has merged into the continuum. In the examples,
$\epsilon=10^{-19}\;\rm{GeV}^{-1}$, $\mu=1$ TeV and $c=0.3$. We
can see how the continuum peaks to higher momenta as the SUSY
breaking mass $m$ increases, specially as the pole merges into the
continuum. }
\label{continuum3}
\end{figure}

For $0<c<1/2$, as we increase $m^2$, the pole eventually merge into the continuum and only a continuum superpartner is left (Fig.~\ref{merge}).
The continuum parts of the spectral functions for several choices of $m^2$ are shown in Fig.~\ref{continuum3}.
\clearpage
%

\begin{figure}[h]
\begin{center}
\epsfig{file=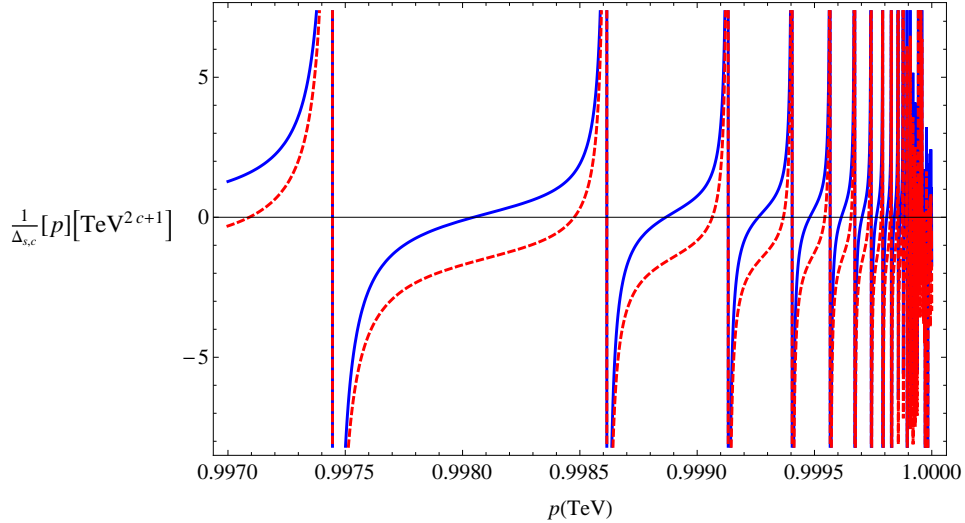,width=5.0in}
\end{center}
%
\caption{Inverse correlator for two examples of boundary UV SUSY
breaking masses: $m=10^{13}$ GeV (blue curve) and $m=2\times
10^{14}$ GeV (red curve, dashed). In the examples,
$\epsilon=10^{-19}\;\rm{GeV}^{-1}$, $\mu=1$ TeV and $c=-0.2$. We
can see how the series of poles shift into the continuum with
increasing values of m. }
\label{SUSYbreaking}
\end{figure}

For $-1/2 < c <0$, the discrete poles below the gap move towards the gap as $m^2$ increase (Fig.~\ref{SUSYbreaking}). However, they do not merge into the continuum for arbitrarily large $m^2$.

\subsection{Gauge Fields}

A 5D $N=1$ vector supermultiplet can be decomposed into a 4D $N=1$ vector supermultiplet $V= (A_\mu, \lambda_1, D)$ and a chiral supermultiplet $\sigma = ((\Sigma + i A_5)/ \sqrt{2}, \lambda_2, F_\sigma)$. We cannot proceed as before by introducing a bulk mass term for the gauge field because of the gauge invariance. The same effect can be obtained with a dilaton profile $\langle \Phi \rangle = e^{-2uz}/ g_5^2$ coupling to the gauge kinetic term, which softly breaks the conformal symmetry in the IR~\cite{Falkowski:2008yr}. The action for the 5D $N=1$ vector supermultiplet can be written as
\begin{eqnarray}
 S_V = 
\int {d^4 xdz \cdot \frac{R}{z}\left\{ {\frac{1}{4}\int {d^2
\theta }  W_\alpha  W^\alpha \Phi +h.c. + \frac{1}{2}\int {d^4
\theta  \left( {\partial _z V - \frac{R}{z}\frac{{(\sigma  +
\sigma ^\dag  )}}{{\sqrt 2 }}} \right)^2 } } \left(\Phi +
\Phi^\dag\right)\right\}}\, .\nonumber\\
\end{eqnarray}
We can obtain the bulk action in components after rescaling the fields, $A_5 \to \frac{z}{R} A_5$, $\lambda_1 \to (\frac{R}{z})^{3/2} \lambda_1$, and $\lambda_2 \to i (\frac{R}{z})^{1/2} \lambda_2$. We find a $z$-dependent bulk gaugino mass $1/2 +uz$ induced by the dilaton profile. It leads to a continuum with a mass gap as we found for the matter fields, and $c$ is fixed to be equal to $1/2$ in this case because the 4D gauge field must have dimension one by gauge invariance.

Adding a SUSY-breaking gaugino mass term on the UV brane will lift the gaugino zero mode. For small Majorana gaugino mass $m$ on the UV brane, the zero mode pole is shifted to
\begin{equation}
p_{\rm pole}^2 \approx \frac{m^2}{(\gamma_E + \ln (2u\epsilon))^2}.
\end{equation}
For large gaugino mass the zero mode will merge into the continuum just as what we saw for the scalar superpartners of matter fields with $c>0$.

\section{Phenomenology and Conclusions}

We discussed a novel possibility for the supersymmetric extension of the Standard Model, where there are continuum excitations of the SM particles and their superpartners arising from conformal dynamics. The properties of the superpartners can be quite different in this type of models. The superpartner of a Standard Model particle can be either a discrete mode below a continuum, or the first of a series of discrete modes, or just a continuum. The continuum superpartners will be quite challenging experimentally. It would be difficult to reconstruct any peak or edge because of the additional smearing of the mass by the continuous spectrum.

\begin{figure}[t]
\begin{center}
\epsfig{file=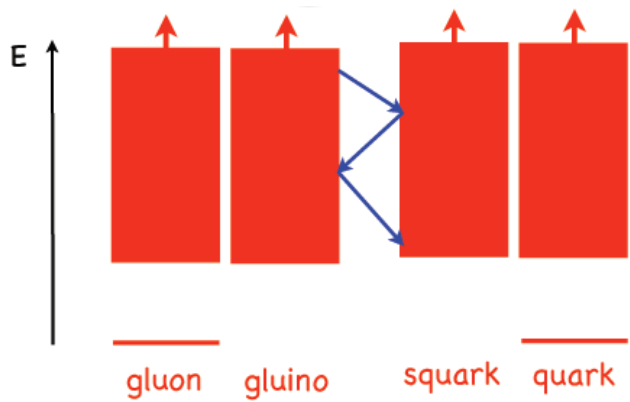,width=3.5in}
\end{center}
\caption{Extened decay chains between the gluino and squark continua.}
\label{chain}
\end{figure}

At a high energy collider, if the superpartners are produced well above the threshold of the continuum, there is a possibility of extended decay chains as shown in Fig.~\ref{chain}.
These events are expected to have large multiplicities and more spherical shapes as a reflection of the underlying conformal theory~\cite{Polchinski:2002jw,Hofman:2008ar,Hatta:2008tx,Csaki:2008dt}.

The experimental searches and verifications of this scenario will be very challenging. It is a topic under current investigation. The LHC will be the only high energy collider in the foreseeable future. We need to be prepared for any surprises and challenges that new physics may present to us at the LHC.

\section*{Acknowledgments}
I would like to thank the organizers for inviting me to give a talk at the SCGT 09 workshop.
This work is supported in part by the Department of Energy Grant DE-FG02-91ER40674.

\end{document}